\documentstyle[preprint,prb,aps]{revtex}
\begin{document}
\draft
\title{Magnetoresistance of a 2-dimensional electron gas \\
in a random magnetic field}
\author{Anders Smith, Rafael Taboryski\cite{raffi}, Luise
Theil Hansen, \\
Claus B. S\o rensen, Per Hedeg\aa rd and  P. E. Lindelof,}
\address{\O rsted Laboratory, Niels Bohr Institute, Universitetsparken 5\\
DK-2100 Copenhagen \O, Denmark}
\date{\today}
\maketitle
\begin{abstract}
We report magnetoresistance measurements on a two-dimensional
electron gas (2DEG) made from a high mobility GaAs/AlGaAs
heterostructure, where the externally applied magnetic field was
expelled from regions of the semiconductor by means of
superconducting lead grains randomly distributed on the surface of the
sample. A theoretical explanation in excellent agreement with the
experiment is given within the framework of the semiclassical
Boltzmann equation.
\end{abstract}

The response of a 2DEG to a spatially inhomogenous magnetic field is  a
subject  of considerable interest both theoretically and
experimentally.\cite{peters}
One of the experimental techniques proposed is
to deposit a pattern of small magnets
on the semiconductor containing the 2DEG.\cite{lead} This technique, however,
only gives rise to a very weak modulation. Another method is to grow
the 2DEG on a substrate with a modulated thickness.\cite{foden} The applied
magnetic field experienced by the electrons in the plane of the curved
2DEG will vary with the thickness modulation. The feasibility of this
method is however limited by the technological difficulties of the
MBE regrowth techniques required. So far, the most simple
technique, originally proposed by Rammer and Shelankov \cite{rammer} for
studying weak localisation effects in inhomogenous magnetic fields,
employs a type II superconducting gate on top of the heterostructure
containing the 2DEG. For a type II superconducting gate an applied
magnetic field  will create the so-called mixed state in the
superconductor above the lower critical field $B_{c1}$. In
this state the magnetic field penetrates the film as
flux tubes. Each flux tube will contain
an integral number of (superconductivity) flux quanta $\Phi_0 =h/2e$.
For a perfect
type II superconductor  the mixed state is accomplished by the
formation of  a two-dimensional hexagonal lattice of vortices. In a
real superconductor inhomogeneities will tend to pin the vortices, so
a random distribution of flux tubes is more likely to occur rather
than the regular lattice. The
magnetoresistance of the type II superconductor gated samples have been
investigated experimentally in various limits of 2DEG properties.
Bending et al. \cite{bend} and Geim\cite{geim} have studied  the weak
localisation effects predicted by Rammer and Shelankov \cite{rammer} for a low
mobility GaAs/AlGaAs heterostructure with a Pb gate and a thin Bi film
evaporated on a Nb/Mo substrate respectively. Kruithof et al. \cite{kruit} have
studied the mechanisms of voltage induction in the 2DEG of a Si MOSFET
caused by flux flow in a Nb/Mo superconducting gate. The above
experiments have probed the diffusive properties of the 2DEG's. In
the ballistic regime, where the electronic mean free path is much
longer than the vortex diameter, a series of experiments were
performed by Geim et al. \cite{geim2,geim3,geim4},
and the results interpreted by
treating the vortices as scatterers. The effect of single vortices
have also been studied both experimentally and
theoretically.\cite{avis,stod,mads}

 In this letter we demonstrate a new and very simple technique for
creating a  strong magnetic field modulation. We also propose a
semiclassical model based on the Boltzmann equation to explain the
measured magnetoresistance caused by the inhomogeneous magnetic
field. Our model gives excellent agreement with the experiment, and in
addition is applicable to the results of Geim et al. \cite{geim4}

 The inhomogeneous magnetic field was achieved by means of small
lead grains randomly distributed on the surface of a high mobility
GaAs/AlGaAs heterostructure. The lead grains (approximated as spheres)
used had a size (average diameter)
distribution as shown in Fig.~\ 1.
For these grain sizes Pb is a type I
superconductor. Below the critical
field given by
\begin{equation}
B_c(T)=B_{c}(0)\left[
1-\left(\frac{T}{T_c}\right)^2 \right], \label{bc}
\end{equation}
there will be (partial) flux expulsion from the grains, creating an
inhomogeneous magnetic field in the 2DEG. Below $\frac{2}{3} B_c(T)$ the
grains will no longer be in the intermediate state, and they will exhibit
full Meissner effect.
For lead $B_{c}(0)= 80.3$ mT. The $T$-dependence in (\ref{bc})
holds to a good approximation for our purpose with $T_c=7.2$ K.
The mobility of the investigated samples was 93.5 m$^2$/Vs
at the lowest temperature of 0.3 K in the
experiment. At 7.2 K the mobility had degraded to 82.6 m$^2$/Vs.
In the whole temperature range the carrier density of the
2DEG was $4.0\times 10^{15}$ m$^{-2}$. This corresponded to a mean free path
of $l \approx 9 \ \mu$m. The experiment was thus performed in a regime
where the mean free path was comparable to the typical grain size
$l_{\rm Pb}$ as well as the average distance between grains
$n_{\rm Pb}^{-1/2}$,
i.\ e.\ $l \leq l_{\rm Pb}, n_{\rm Pb}^{-1/2}$, with $n_{\rm Pb}$  being
the density of lead grains. The sample geometry consisted of a
standard $400 \ \mu$m wide Hall bar with 3 pairs of voltage
probes, each pair placed on opposite sides of the Hall bar. The
voltage probe pairs were  displaced a distance of $1600 \ \mu$m
(4 squares) from each other. In addition the Hall bar also contained
two current probes displaced another 4 squares from the voltage
probes. The resistance was measured by conventional small signal
lock-in techniques in a current controlled four-probe configuration.
To check the homogeneity of the lead grain distribution we measured
the longitudinal magnetoresistance between all three combinations of
voltage probes along each side of the Hall bar.  Such measurements
always gave the same result within 5\% when normalised with respect to
the number of squares between the voltage probes. We also made control
measurements on samples cut from the same heterostructure, but without
lead grains. Such samples showed no magnetoresistance in fields below
0.1 T.  The measurements were performed in the following
way. First, we cooled down the sample  in zero magnetic field to the
relevant temperature. Then we swept the magnetic field to above the
critical field while measuring the magnetoresistance. This was
followed by consecutive down and up sweeps as exemplified in Fig.~\ 3.
Here we should emphasize that the first sweep after each cool down
procedure is fundamentally different from the following sweeps. This
difference is caused by the trapping of flux in the superconductor. In
fact, we believe that the magnetoresistance in the consecutive sweeps
is dominated by the random magnetic field caused by the frozen flux.
We exploit both experimental situations to test our theoretical model
for two different realisations of a random magnetic field.

 In Fig.~\ 2 we show a set of `sweep up' traces at different
temperatures. The pronounced magnetoresistance peak is seen to vary in
amplitude and width with temperature. Moreover, the peaks are
asymmetric in the field. However, as seen in Fig.~\ 3 the corresponding
`sweep down' traces are asymmetric as well but with the maximum
resistance at negative magnetic fields. As indicated by the vertical
dashed lines in Fig.~\ 2 the magnetoresistance
defined as $\Delta\rho_{xx}(B)=\rho_{xx}(B)-\rho_0$
goes to zero at $B=B_c(T)$. Here $\rho_0$ is the resistance
at zero magnetic field prior to the first sweep
after the cool down procedure. It is also seen from Fig.~\ 2 that the
magnetoresistance peak vanishes for temperatures above the critical
temperature of lead ($T_c=7.2$ K). The observed
magnetoresistance shown in figures 2 and 3
cannot originate from  any
weak localisation contribution to the magnetoresistance. The weak
localisation magnetoresistance is extremely weak for high mobility
GaAs/AlGaAs samples and is practically extinguished for magnetic
fields $B \geq 4 B_\phi \simeq 10^{-3}$ T, where $B_\phi$ is the
characteristic magnetic field corresponding to the phase breaking
scattering time.\cite{tab} Weak localisation effects will in addition not
show the hysteresis effect displayed in Fig.~\ 3.
Moreover the observed
magnetoresistance is fundamentally different from the curves reported
in Ref.\ \onlinecite{geim4} with a continuous lead gate.
This difference is most
easily seen on the `first sweep' curve in Fig.~\ 3, which for small
magnetic fields has a $\Delta\rho_{xx} \propto B^2$  dependence, while the
magnetoresistance observed in Ref.\ \onlinecite{geim4} exhibits a
$\Delta\rho_{xx} \propto B$ dependence in the same regime of fields.

 We now turn to describe the theory. Disregarding
interference effects and wavevector quantization imposed by sample
boundaries, one can in general expect a classical approach to
conduction in a magnetic field to be valid if $k_F l \gg 1$,
where $k_F$ is the Fermi wave vector and $l$
is the electronic mean free path. A randomly modulated magnetic
field can be included in the Boltzmann equation either in the driving
force term or as an effective impurity cross section, depending on the
correlation length $a$ of the modulation (in our case given by the
size of the lead grains).  If
$a \gg 1/k_F$,
as definitely is the case in our experiment, a modulation
$\delta B$ of the magnetic field can be treated as an ordinary
external field in the driving force term of the Boltzmann equation.
However, if $a \simeq 1/k_F$, $\delta B$ should be
incorporated as an impurity cross section. (When $a\gg 1/k_F$ we could
of course also treat $\delta B$
as a scatterer, i.\ e.\  put it on the RHS of the Boltzmann equation.
By contrast it would be
inconsistent to put $\delta B$ on the LHS when $a\sim 1/k_F$.)

 Our starting point is thus the usual semiclassical Boltzmann equation
in the relaxation time approximation. We introduce polar co-ordinates
$v,\phi$ for the velocity and confine ourselves to $T=0$.
Then $v$ only enters through a delta-function
and can be put equal to $v_F$. Furthermore we will make
the usual  assumption of a constant, external driving field $\bbox{E}$
and calculate to linear order in it. The resulting
equation is
\begin{equation}
\left\{ v_F \left( \begin{array}{c} \cos\phi \\ \sin\phi \end{array}
\right)\cdot
\frac{\partial }{\partial\bbox{r}} + \omega_c(\bbox{r})
\frac{\partial}{\partial\phi}
+\frac{1}{\tau}\right\}g(\bbox{r},\phi) = -\frac{e}{m}\bbox{E}\cdot
\left( \begin{array}{c} \cos\phi \\ \sin\phi \end{array}
\right) \delta(v-v_F) .
\end{equation}
Here $\omega_c(\bbox{r})=eB(\bbox{r})/m$ is a function of position.
Writing $\omega_c(\bbox{r})=\omega_0 + \delta\omega(\bbox{r})$, defined such
that $<\delta\omega>=0$ ($<>$ denotes an average over random magnetic
field configurations)
we can write the Boltzmann equation as an operator
equation
\begin{equation}
Dg \equiv (D_0 + W)g = \chi,
\end{equation}
with $W=i \delta\omega(\bbox{r})\partial/\partial\phi$
(we have
multiplied the equation with $i$ for convenience). The Green's function
for $D_0$ is readily found, and calculating to second
order using the Dyson equation
\begin{equation}
D^{-1} \simeq D_0^{-1} + D_0^{-1}<WD_0^{-1}W>D^{-1}
\end{equation}
we find \cite{smith} the
resistivity tensor
\begin{equation}
\bbox{\rho} = \frac{m}{ne^2\tilde{\tau}}
 \left( \begin{array}{cc} 1 & \tilde{\omega}\tilde{\tau} \\
                         -\tilde{\omega}\tilde{\tau} & 1 \end{array}\right),
\end{equation}
where we have defined the renormalised quantities
\begin{eqnarray}
\tilde{\omega} & = & \omega_0 - \mbox{Re}\, \Sigma_1 \\
\frac{1}{\tilde{\tau}} & = & \frac{1}{\tau} - \mbox{Im}\, \Sigma_1,
\end{eqnarray}
with the `self-energy' $\Sigma_1$ given by
\begin{equation}
\Sigma_1 = -\frac{i}{\omega_0\pi}\frac{1}{e^{2\pi/\omega_0\tau}-1}
\int_{0}^{2\pi}f(2r_c\sin\frac{\theta}{2})e^{(i+1/\omega_0\tau)\theta}d\theta.
\label{self}
\end{equation}
Here $r_c=v_F/\omega_0$ is the average cyclotron radius and
$f(r)=<\delta\omega(\bbox{r})\delta\omega(0)>$ is the correlation function,
depending on the nature of the random magnetic field modulation.
We see that the change in $\bbox{\rho}$ is directly related to $\Sigma_1$:
\begin{eqnarray}
\frac{\Delta\rho_{xx}}{\rho_{xx0}} & = & -\tau \mbox{Im}\, \Sigma_1 \\
\frac{\Delta\rho_{xy}}{\rho_{xy0}} & = & -\omega_0^{-1} \mbox{Re}\, \Sigma_1.
\end{eqnarray}

To proceed we now introduce a model for the
modulated magnetic field. We start by treating the situation with
perfect flux expulsion due to the Meissner effect applicable to the
`first sweep' curves. The lead grains had a distribution of sizes as
seen in Fig.\ 1. We represent this distribution by an average size
$l_{\rm Pb}$. We model the magnetic field modulation
from a single lead grain by\cite{corr}
$\delta b(r)=B((r/l_{Pb})^2-1)e^{-r^2/l_{\rm Pb}^2}$.
This expression fulfils the necessary
flux conservation condition $\int dr\, r\, \delta b(r) = 0$.
The magnetic field
modulation can now be expressed by $\delta B(\bbox{r}) =
\sum_i \delta b(\bbox{r} - \bbox{R}_i)$, where $\bbox{R}_i$
is the position of the i'th grain. $\bbox{R}_i$ is
randomly distributed,
and the magnetic field modulation should be averaged over
different distributions of lead grains on the surface of the
semiconductor. This gives rise to the correlation function
\begin{equation}
f(r)=B^2\frac{n_{\rm Pb}l_{\rm Pb}\pi}{32}\left\{ 8 -
8\left(\frac{r}{l_{\rm Pb}}\right)^2 +
\left(\frac{r}{l_{\rm Pb}}\right)^4 \right\}
e^{-\frac{r^2}{l_{\rm Pb}^2}}.  \label{corr}
\end{equation}
At low magnetic fields we get $\Delta\rho_{xx} \propto B^2$
in accordance
with the experiment.
At higher fields ($> \frac{2}{3} B_c$ for spheres) the grains will be
in the intermediate state, reducing the amplitude of $f$ until it
vanishes at $B=B_c$. This we model by multiplying the correlation
function in (\ref{corr})  by a factor going to zero as
$1-\sqrt{\frac{B}{B_c(T)}}$
when $B \rightarrow B_c$.
The temperature dependence of $B_c(T)$ accounts for the
observed temperature dependence of the magnetoresistance as shown in
Fig.~\ 2.

In the case when the magnetic modulation is caused by frozen
flux, the correlation function (\ref{corr}) should be
replaced by
\begin{equation}
f(r)=C(B) e^{-\frac{r^2}{l_{\rm Pb}^2}}, \label{frozcorr}
\end{equation}
 where $C(B)$ is an asymmetric function of $B$ going to zero for
$|B|>B_c$, giving a phenomenological measure of the amount of flux
trapped in the lead grains. The position of the maximum of $C$ is used
as a fitting parameter.
In Fig. 3 we have shown a fit of the model to
the experimental traces. The correspondence is seen to be quite
satisfactory.

 Finally we demonstrate that our theoretical model is applicable to
the experiments in Ref. \onlinecite{geim2}, where the magnetic modulation was
produced by filaments of magnetic field emerging from a type II
superconducting gate. Each fluxtube may be modelled by the following
expression\cite{tink} $b(r)=(\Phi_0/2\pi \lambda_L^2)K_0(r/\lambda_L)$,
where $\lambda_L$ is the
effective London length in the plane of the 2DEG and
$K_0$ a modified
Bessel function. In this case the correlation function turns out to
be:
\begin{equation}
f(r)=n_{\nu} \frac{\Phi_0^2}{4\pi \lambda_L^2}\frac{r}{\lambda_L}
K_{-1}\left(\frac{r}{\lambda_L}\right) , \label{vortcorr}
\end{equation}
where $n_\nu$ is the density
of vortices.
Since $n_\nu \propto B$
it is immediately seen that the longitudinal magnetoresistance will be
proportional to $B$ for small fields, as is also
found in Ref.\ \onlinecite{geim2}. We can also find the density
dependence of $\Delta\rho_{xx}$ within this framework, and find that
$\Delta\rho_{xx} \propto n^{-3/2}$. However, when Eq.(\ref{vortcorr})
is used to fit the
experimental traces with $\lambda_L$ as the
fitting parameter, the obtained values of $\lambda_L$ are
approximately a factor of 5--10 larger than estimated in
Ref.\ \onlinecite{geim2}.
This may be a result of flux-bundles containing several
flux-tubes, as also reported by Stoddart et al. \cite{stod}  We would like to
emphasize both that our calculation is purely semiclassical, and that
we do not
treat the vortices as scatterers to be included in a scattering cross
section, but rather as a perturbation to the driving force term in the
Boltzmann equation.

 In conclusion we have measured the magnetoresistance of a 2DEG
subject to a random shielding of the externally applied magnetic
field. We have modelled our results by solving the semiclassical
Boltzmannn equation with an appropriately chosen random magnetic field
in the sample. We have demonstrated that our model can be applied in
electrical transport problems with  other types of magnetic field
modulation.

 We would like to thank Dr Andrey Geim, Dr Crispin Barnes, Dr Clare
Foden, Mr Boaz Brosh, Mads Nielsen and Karsten Juel Eriksen
for valuable discussions.

\begin{figure}
\caption{Distribution of  lead grain sizes on the sample for which
data is presented in this paper. This distribution was simply obtained
by measuring the grain sizes on the sample on a photo taken through an
optical microscope.}
\end{figure}
\begin{figure}
\caption{A series of  sweep up curves at different
temperatures, displaced for clarity. The temperatures were
$0.3$ K, $1.3$ K, $4.4$ K, $6.0$ K, $7.2$ K and, $8.5$ K, where
the upper curves corresponds to the lowest temperatures.
The magnetoresistance anomaly disappears when the sample
is heated to above the critical temperature for lead ($T_c= 7.2$ K).
The vertical dashed lines indicate the critical magnetic field
calculated with equation (1).}
\end{figure}
\begin{figure}
\caption{A set of  magnetoresistance traces taken at 4.4 K
 (solid lines). The `first sweep below $T_c$' curve is
taken just after the sample was cooled from above the critical
temperature in zero magnetic field. The `sweep down'  and `sweep up'
curves are the subsequent sweeps, after the first sweep where the
magnetic field was taken to above the critical field. The  dashed
curves are calculated with equations
(\protect{\ref{self}}--\protect{\ref{frozcorr}}) with
$l_{\rm Pb}= 11.5\ \mu$m and $n_{\rm Pb}= 7.75\  10^8\ $m$^{-2}$.
The parameters are taken from the distribution of lead grain sizes
shown in fig. 1, and the position of the maximum (used as a fitting
parameter) is consistent with
the trapped flux interpretation. The vertical dashed line indicate the critical
magnetic field at the relevant temperature.}
\end{figure}

\begin{references}
\bibitem{corr} The exact form of $\delta B$ is not
important, only the prefactor of $B$ and the fact that there is a
fixed number of grains (in contrast to the vortex case where $n_\nu
\propto B$).
\end{references}
\end{document}